# Enhancing Cybersecurity in IoT Networks: A Deep Learning Approach to Anomaly Detection

Yining Pang
Hebei Finance University
Baoding, China
p18132109069@163.com

Chenghan Li*
ZJU-UIUC Institute, Zhejiang University
Haining, China
chenghan.20@intl.zju.edu.cn

*Abstract*—With the proliferation of the Internet and smart devices, IoT technology has seen significant advancements and has become an integral component of smart homes, urban security, smart logistics, and other sectors. IoT facilitates real-time monitoring of critical production indicators, enabling businesses to detect potential quality issues, anticipate equipment malfunctions, and refine processes, thereby minimizing losses and reducing costs. Furthermore, IoT enhances real-time asset tracking, optimizing asset utilization and management. However, the expansion of IoT has also led to a rise in cybercrimes, with devices increasingly serving as vectors for malicious attacks. As the number of IoT devices grows, there is an urgent need for robust network security measures to counter these escalating threats. This paper introduces a deep learning model incorporating LSTM and attention mechanisms, a pivotal strategy in combating cybercrime in IoT networks. Our experiments, conducted on datasets including IoT-23, BoT-IoT, IoT network intrusion, MQTT, and MQTTset, demonstrate that our proposed method outperforms existing baselines.

## I. Introduction

The Internet of Things (IoT) represents the third revolution within the information technology sector, significantly contributing to economic growth and human advancement. [1] The proliferation of smart devices across diverse domains has led to a burgeoning integration of these devices into the IoT network. Projections indicate that by 2025, the global IoT device count will soar to 25.2 billion, encompassing a spectrum of applications, including smart grids, smart meters, smart home appliances, telematics, and asset tracking. [2]

As an emerging technology, the Internet of Things (IoT) heralds the era of the Internet of Everything, enhancing convenience in daily life through its integration with sectors such as healthcare, education, and travel. This integration fosters the emergence of new models that cater to the public's evolving, diversified, and quality-driven needs. Nonetheless, it also heightens concerns regarding network security and privacy protection. IoT facilitates intelligent functions such as identification, positioning, tracking, and monitoring through the analysis of vast datasets. [3]However, the diversity of IoT devices' hardware and software ecosystems presents a broad attack surface, which, coupled with the proliferation of IoT devices, provides attackers with an increased number of potential targets and opportunities for data theft. Consequently, there is an elevated demand for robust network security management, with a particular emphasis on the anomaly detection of IoT devices.

In the digital transformation wave, network security is pivotal to the growth of the digital economy. Ensuring the broad application of digital technologies hinges on robust network security, which is becoming increasingly critical. The timely and efficient detection and classification of malicious attacks represent a pressing challenge. Machine learning and deep learning are instrumental in identifying malicious IP addresses. These technologies enable data-driven learning algorithms to discern features and patterns, distinguishing between benign and malicious traffic. They facilitate intelligent and reliable anomaly detection and unauthorized access prevention, thereby enhancing the precision and efficiency of network monitoring and malicious node identification. To optimize the runtime and power consumption on mobile devices, we advocate the use of lightweight machine learning algorithms and neural networks, which also aim to improve the accuracy of malicious node detection. A central unit within the model captures IoT traffic data, subsequently directing it to a chosen, pre-trained machine learning or deep learning model. To cater to the distinct requirements of various users or groups, a variety of trained models must be evaluated to identify the most efficient solutions for different user profiles.[4][5]

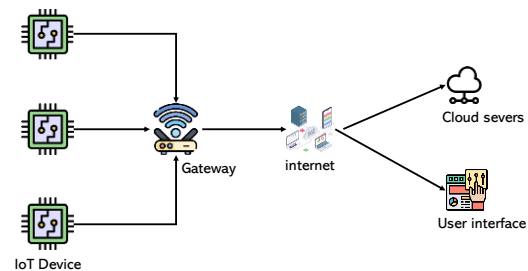

Figure 1. The architecture of a typical IoT deployment

Following an attack, Internet of Things (IoT) devices often serve as conduits for assailants to further target other IoT devices or servers, potentially leading to extensive network disruptions. The constrained computational capabilities and storage limitations of IoT devices render the application of traditional cybersecurity measures, such as firewalls and antivirus programs, impractical. Moreover, the vast diversity within the hardware and software ecosystems of IoT devices introduces significant challenges. The heterogeneity inherent in IoT environments results in substantial functional and traffic pattern variations among devices. Existing anomaly detection models typically lack the nuanced customization necessary to address these differences, thereby complicating the achievement of cost-effectiveness, reliability, and performance objectives. Furthermore, given the resource constraints of contemporary IoT devices, they must be capable of detecting intrusions with minimal complexity and expedited timing. The integration of machine learning and deep learning technologies can attenuate the computational demands for intrusion identification and detection. This paper is structured as follows: Section II provides a literature review on IoT networks, Section III outlines the methodology of this study, Section IV presents data analysis using the IoT-23 dataset along with results, and Section V concludes with a discussion of findings and future research directions.

## II. LITERATURE REREVIEW

Machine learning-based IoT inspection techniques extract statistical features from network traffic to apply machine learning or deep learning algorithms for classification. This section will introduce some algorithms and methods for detecting anomalies, involving various mechanisms to improve the accuracy and efficiency of identifying malicious attacks on IoT devices. For example, Arunan Sivanathan et al. [6] simulated a smart environment using 28 IoT devices, including cameras, plugs, motion sensors, appliances, etc., and collected traffic traces for 6 months using data from flow volume, flow duration, average flow rate, device sleep time, server port number, DNS queries, NTP queries, and cipher suite. A classification framework based on multi-stage machine learning is developed. The framework achieved a comprehensive accuracy of more than 99%. Vanhoenshoven et al. [7] studied the performance of several mainstream classifiers in machine learning, namely naive Bayes, support vector machine, multi-layer perceptron, decision tree, random forest, and K-proximity algorithm, conducted classification experiments on a large number of malicious URLs, and numerical simulation showed that random forest and multi-layer perceptron achieved the highest precision and accuracy. Dutta et al. [8]used the deep autoencoder to reduce the dimensionality of the Zeek log data samples in the IoT-23 dataset and then used LSTM for classification. Finally, after comparing the test results with random forest, SVM, MLP, and other algorithms, they found that this method improved the classification performance. In [9], a convolutional neural network-based intrusion detection model for the Internet of Things is designed and developed, which converts feature vectors into one-dimensional, two-dimensional, and three-dimensional shapes and finally achieves 99.9% high accuracy, accuracy, recall rate, and F1-score by using corresponding convolutional neural networks for classification. According to[10], a mechanism using random time-jump sequences and permutations to hide verification information is proposed. This mechanism detects data tampering in IoT systems through efficient and simple techniques. Performance analysis shows that it has low computational complexity and is suitable for IoT systems. Yin et al. [11] applied deep learning technology to propose a Bot-GAN model based on a Generative Adversarial Network framework to enhance botnet detection (BoT-GAN), which includes a generative adversarial network (GAN) to generate simulation data to enhance the original data. The experimental results show that Bot-GAN can improve the detection performance and reduce the false positive rate, which is suitable for enhancing the original detection model. Khan and Cotton [12] introduced a feature-free machine learning process for anomaly detection, using unprocessed packet byte streams as training data. This method effectively shortened the test time, and the test results on the IoT-23 dataset showed that the detection accuracy reached 100%. Supports low-cost and low-memory time series analysis of network traffic. Ullah and Mahmoud [13] jointly developed an anomaly recognition model for the Internet of Things based on RNN.

Upon review, we identify several key issues with existing IoT anomaly detection methodologies: previous research models have uniformly treated various IoT feature types without tailoring network architectures to specific feature categories. Furthermore, while achieving high detection rates, these studies predominantly concentrate on binary classification, discerning the presence of anomalies but not their specific nature. In response to these issues, our contributions are delineated as follows:

- ✧ A multi-variable feature extraction model architecture based on LSTM and Attention mechanisms is proposed.
- ✧ The effectiveness of this architecture has been validated on datasets including IoT-DS2, NSLKDD, IoT-23, MQTT, IoT-NI, BoT-IoT, and MQTTset.
- ✧ Further, ablation and comparison experiments have been conducted to validate the effectiveness of each component of the proposed architecture.

## III. METHODOLOGY

The proposed model architecture is an advanced neural network tailored for sequence processing, featuring Long Short-Term Memory (LSTM) units and an integrated attention mechanism. Initially, the model deploys two distinct LSTM layers to handle separate sequence inputs. These layers are pivotal in capturing the sequences' temporal dependencies, enabling the model to comprehend the sequential data characteristics.(EcNet) Subsequently, the model integrates an attention mechanism, which selectively emphasizes the significance of various input sequence segments, facilitating the model's focus on pertinent information for prediction. The attention-weighted outputs are subsequently channeled into fully connected (FC) layers, refining these outputs into an optimal format for the final predictive task. The FC layers' transformation ensures the model's capacity to discern complex data representations. Ultimately, the model's output undergoes processing by a softmax activation function, which translates it into a probability distribution. Figure 2 illustrates the structure of the proposed model.

### A. LSTM

The internal architecture of the Long Short-Term Memory (LSTM) unit is depicted in Figure 3, accompanied by the corresponding mathematical formulation presented below:

$$f_t = \sigma(W_\mathrm{f} \cdot [h_{t-1}, x_t] + b_\mathrm{f}) \quad (1)$$

$$i_t = \sigma(i_\mathrm{f} \cdot [h_{t-1}, x_t] + b_\mathrm{i}) \quad (2)$$

$$g_t = \tanh(W_c \cdot [h_{t-1}, x_t] + b_c) \quad (3)$$

$$o_t = \sigma(W_\mathrm{o} \cdot [h_{t-1}, x_t] + b_\mathrm{o}) \quad (4)$$

$$c_t = f_t \otimes c_{t-1} + i_t \otimes g_t \quad (5)$$

In the LSTM formula, $f_t$ is the output of the forgetting gate; $h_{t-1}$ and $x_t$ are the inputs into the sigmoid activation function ($\sigma$). It is the output of the input gate that controls the deletion or retention of the candidate memory unit $g_t$. $o_t$ is an output gate. $W_\mathrm{o}$ is the corresponding weight parameter; $b_\mathrm{o}$ is the corresponding offset.

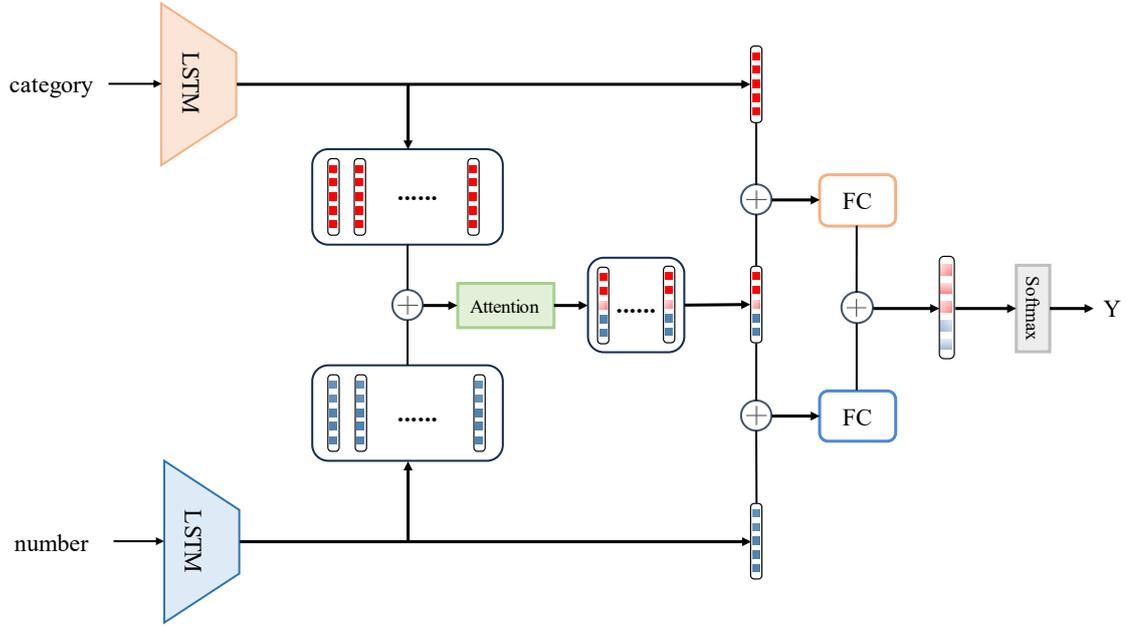

Figure 2. The structure of our proposed model.

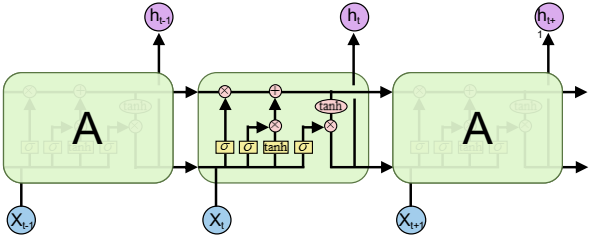

Figure 3. The structure of LSTM

### B. Attention

The output vectors derived from the Long Short-Term Memory (LSTM) units undergo three distinct mapping operations, transforming the vector into three input matrices: Query $Q$, Key $K$, and Value $V$, each with dimension $d_k$. The attention output matrix is delineated in the accompanying equation and depicted in Figure 4.

$$\text{Attention}(Q, K, V) = \text{softmax}\left(\frac{QK^T}{\sqrt{d_k}}\right)V$$

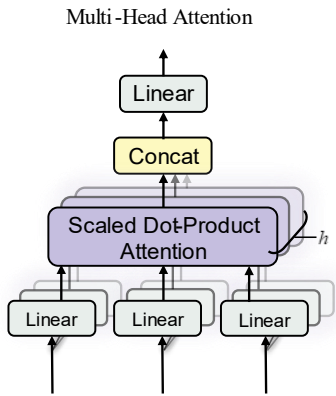

Figure 4. The Structure of Multi-head Attention

## IV. EXPERIMENTS

This section introduces our proposed methodology for detecting anomalies within IoT networks. A range of metrics is utilized to assess the performance of our approach, and a comparative analysis is conducted with established machine learning (ML) and deep learning (DL) baseline models.

### A. Datasets

*IoT-23*: The dataset utilized in this study originates from the IoT-23 dataset, published in January 2020 [14], encompassing network traffic data from devices such as Amazon Echo, Philips Hue, and Somfy Door Lock. It comprises 23 network captures, with 20 classified as malicious and 3 as benign. Each capture from a compromised device is annotated with the specific malware executed, including categories such as Attack, Command and Control (C&C), Distributed Denial of Service (DDoS), and Okiru. The dataset is presented in the Zeek conn.log.labeled format, which is derived from the original pcap files using the Zeek network analyzer. We implemented a sampling strategy to manage dataset size to select records that preserve most attack types from the original IoT-23 dataset. Table 1 delineates the combined dataset, which encompasses 10 distinct attack types, namely: PartOfAHorizontalPortScan, Okiru, DDoS, Attack, C&C-HeartBeat, C&C FileDownload, C&C-Torii, FileDownload, C&C-HeartBeat FileDownload, and C&C-Mirai.

TABLE I.  LABEL COUNTS OF IOT-23

| Label | count |
|---|---|
| PartOfAHorizontalPortScan | 825939 |
| Okiru | 262690 |
| Benign | 197809 |
| DDoS | 138777 |
| Attack | 3915 |
| C&C-HeartBeat | 349 |
| C&C-FileDownload | 43 |
| C&C-Torii | 30 |
| FileDownload | 13 |
| C&C-HeartBeat-FileDownload | 8 |
| C&C-Mirai | 1 |

*BoT-IoT:* The BoT-IoT dataset, referenced herein [15], encompasses virtual machines (VMs) that interface with the network through both LAN and internet connections, with a PFSense firewall facilitating the VMs' internet access and an Ubuntu server supplying IoT-specific resources. The dataset features normal traffic, which is generated using the Ostinato tool and involves an ensemble of five IoT devices in conjunction with a cloud server. It comprises a total of 35,960,520 instances categorized into five distinct classes: normal, Denial of Service (DoS), Distributed Denial of Service (DDoS), Scan, and Data theft.

TABLE II.  LABEL COUNTS OF BOT-IOT

| Label | count |
|---|---|
| Normal | 9543 |
| TCP-DoS | 12315997 |
| UDP-DoS | 20659491 |
| HTTP-DoS | 29706 |
| TCP-DDoS | 19547603 |
| UDP-DDoS | 18695106 |
| HTTP-DDoS | 19771 |

*IoT network intrusion:* The dataset referenced [16], pertinent to smart home systems equipped with IoT devices, encompasses a suite of integrated devices, including laptops and smartphones, connected through a smart home router to assess device vulnerabilities. It delineates four principal attack categories, each further differentiated into five distinct types.

TABLE III.  LABEL COUNTS OF IOT NETWORK INTRUSION

| Label | count |
|---|---|
| DDos | 92129 |
| DoS | 82509 |
| Reconnaissance | 3331 |
| Normal | 22 |
| Theft | 4 |

*MQTTSet:* The dataset [17] comprises 10 Internet of Things (IoT) sensors within a smart home testbed environment, created utilizing IoT Flock, a network traffic simulation tool, in conjunction with IoT devices.

TABLE IV.  LABEL COUNTS OF MQTTSET

| Label | count |
|---|---|
| Normal | 440699 |
| Bruteforce | 4547 |
| MQTTFlood | 77793 |
| MalariaDos | 11408 |
| Malformed | 3580 |
| SlowITe | 3044 |

*MQTT:* The dataset [18] encompasses brute force attack scenarios and incorporates 12 MQTT sensors. It simulates the system that captures network characteristics from a camera to detect intruders. Concurrently, these sensors provide a comprehensive display of network traffic details. The MQTT dataset has been authenticated against real-world application scenarios.

TABLE V.  LABEL COUNTS OF MQTT

| Label | count |
|---|---|
| MQTT Brute Force attack | 1048563 |
| Aggressive scan attack | 502927 |
| UDO scan attack | 22359 |
| Sparta SSH brute-force attack | 32405 |
| MitiM attack | 7000 |
| Dos | 288657 |
| Buffer overflow attack | 9235 |

### B. Evaluation measures

*Accuracy:* accuracy is defined as the proportion of accurate forecasts relative to the overall count of forecasts made.

$$accuracy = \frac{TP + TN}{TP + FP + TN + FN} \quad (6)$$

*Precision:* precision refers to the proportion of correct positive identifications in relation to all instances classified as positive.

$$precision = \frac{TP}{TP + FP} \quad (7)$$

*Recall:* recall is the metric that quantifies the accuracy of positive predictions by dividing the number of true positives by the total number of predicted positives (which includes both true and false positives).

$$recall = \frac{TP}{TP + FN} \quad (8)$$

*F1-score:* The F1 score is the harmonic mean of the true positive rate (recall) and positive predictive value (precision).

$$F1 = \frac{2 * precision * recall}{precision + recall} \quad (9)$$

In these formulas, true positives are denoted by *TP*, false positives by *FP*, true negatives by *TN*, and false negatives by *FN*.

### C. Result Analysis

Table VI provides a comparative analysis of the performance metrics for various machine learning models, namely SVM, XGBoost, LSTM, and our proposed model, across five datasets: BoT-IoT, IoT-23, IoT-NI, MQTT, and MQTTset. The evaluation criteria included accuracy, precision, recall, and F1-score. The findings reveal that our proposed model outperforms the other models on all datasets. For instance, on the BoT-IoT dataset, it attained the highest accuracy of 88.57%, precision of 91.08%, recall of 89.62%, and an F1-score of 90.31%. Comparable superior performance is noted for the other datasets. In the IoT-23 dataset, the model achieved an accuracy of 86.09%, a precision of 88.57%, a recall of 86.62%, and an F1-score of 87.53%. Although the XGBoost model performed well, it consistently ranked second to our model, especially in precision and recall. The LSTM and SVM models exhibited relatively lower performance. This analysis underscores the efficacy of our proposed model in advancing the performance of machine learning tasks focused on IoT data classification. The metrics are computed in an average manner.

TABLE VI. EXPERIMENTS OF RESULTS IN DATASETS

| Dataset | Model | Accuracy | Precision | Recall | F1-score |
|---|---|---|---|---|---|
| BoT-IoT | SVM | 73.24 | 76.10 | 72.16 | 74.07 |
| | XGBoost | 84.09 | 88.73 | 85.20 | 86.93 |
| | LSTM | 82.15 | 85.30 | 83.50 | 84.39 |
| | Our | **88.57** | **91.08** | **89.62** | **90.31** |
| IoT-23 | SVM | 70.05 | 72.58 | 68.13 | 70.23 |
| | XGBoost | 81.53 | 85.07 | 82.11 | 83.52 |
| | LSTM | 80.04 | 83.07 | 80.56 | 81.78 |
| | Our | **86.09** | **88.57** | **86.62** | **87.53** |
| IoT-NI | SVM | 74.00 | 77.00 | 73.50 | 75.20 |
| | XGBoost | 83.07 | 87.13 | 84.03 | 85.52 |
| | LSTM | 81.03 | 84.07 | 81.59 | 82.78 |
| | Our | **87.54** | **90.09** | **88.03** | **89.03** |
| MQTT | SVM | 72.00 | 74.50 | 71.00 | 72.71 |
| | XGBoost | 82.07 | 86.13 | 83.11 | 84.52 |
| | LSTM | 79.54 | 82.59 | 80.13 | 81.28 |
| | Our | **85.17** | **88.07** | **86.13** | **87.01** |
| MQTTset | SVM | 75.00 | 78.00 | 74.50 | 76.21 |
| | XGBoost | 84.50 | 88.50 | 85.50 | 86.97 |
| | LSTM | 82.03 | 85.07 | 83.11 | 84.01 |
| | Our | **88.03** | **91.07** | **89.11** | **90.00** |

*LSTM vs. RNN vs. GRU:* Figure 5 displays the accuracy performance of three neural network models—LSTM, RNN, and GRU—evaluated across five distinct datasets: BoT-IoT, IoT-23, IoT-NI, MQTT, and MQTTset. Each subplot provides a comparative accuracy analysis of each dataset's models. The LSTM model demonstrates a consistent superiority over the RNN and GRU models in accuracy across all datasets, securing the top performance. Although the RNN and GRU models show slightly diminished accuracy, their performance remains robust, exceeding 70%. These findings indicate that the LSTM model may be the optimal choice for accuracy in the classification of these dataset types.

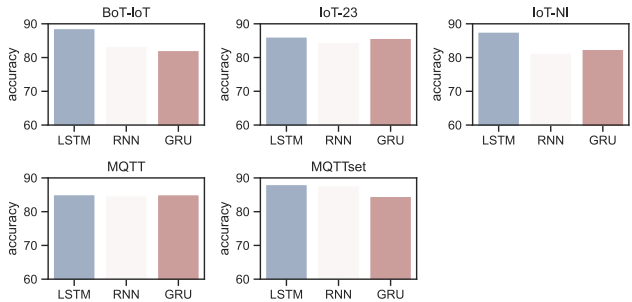

Figure 5. Results of the different module (LSTM vs. RNN vs GRU)

*Attention vs. Non-Attention:* Table VII presents a detailed ablation study that meticulously examines the contribution of the attention mechanism within our model. The study findings underscore the mechanism's remarkable efficacy in identifying and emphasizing the most pertinent features, enhancing the model's predictive accuracy. This discernment of feature significance is a testament to the attention mechanism's ability to filter out less relevant information, focusing instead on the nuances that truly matter for the task. Experiments of results for attention.

TABLE VII. RESULTS OF THE ATTENTION MODULE

| Dataset | Model | Accuracy | Precision | Recall | F1-score |
|---|---|---|---|---|---|
| BoT-IoT | w/o | 84.23 | 87.15 | 85.48 | 86.32 |
| | Our | 88.57 | 91.08 | 89.62 | 90.31 |
| IoT-23 | w/o | 81.74 | 85.21 | 83.67 | 84.23 |
| | Our | 86.09 | 88.57 | 86.62 | 87.53 |
| IoT-NI | w/o | 83.12 | 85.79 | 84.23 | 84.98 |
| | Our | 87.54 | 90.09 | 88.03 | 89.03 |
| MQTT | w/o | 83.40 | 80.91 | 83.45 | 81.98 |
| | Our | 85.17 | 88.07 | 86.13 | 87.01 |
| MQTTset | w/o | 84.56 | 87.09 | 85.23 | 85.97 |
| | Our | 88.03 | 91.07 | 89.11 | 90.00 |

w/o means the EcNet doesn't have Attention module

*Binary Classify:* We further extended our evaluation by conducting binary classification tasks, distinguishing between abnormal and normal states, across five diverse datasets to rigorously test the efficacy of our proposed method. Figure 8 shows the confusion matrix for binary classification in anomaly detection. Figure 8 vividly displays the comparative analysis, illustrating the superior performance of our approach. The results conclusively demonstrate that our method not only excels in this binary classification context but also consistently outperforms existing benchmarks, showcasing its robustness and reliability. This comprehensive evaluation across varied datasets reinforces the effectiveness of our method in accurately identifying anomalies, which is crucial for its practical applicability in real-world scenarios.

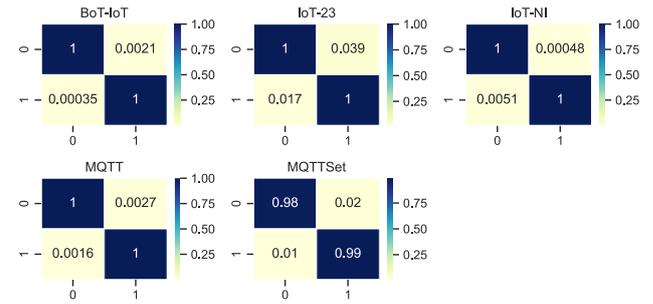

Figure 6. Binary classification results in datasets

Table VIII presents the outcomes of tests conducted to evaluate our distinct feature extraction techniques. The results indicate that, across various neural network architectures, the separate feature extraction approach is effective in most instances.

TABLE VIII. RESULTS OF FEATURE EXTRACTION WAYS

| Dataset | Model | LSTM | RNN | GRU |
|---|---|---|---|---|
| BoT-IoT | ✓ | 88.57 | 83.24 | 82.08 |
| | ✗ | 87.45 | 82.13 | 82.45 |
| IoT-23 | ✓ | 86.09 | 84.51 | 85.67 |
| | ✗ | 86.12 | 83.23 | 84.12 |
| IoT-NI | ✓ | 87.54 | 81.23 | 82.41 |
| | ✗ | 86.32 | 81.67 | 82.45 |
| MQTT | ✓ | 85.17 | 84.69 | 85.01 |
| | ✗ | 84.89 | 83.45 | 85.56 |
| MQTTset | ✓ | 88.03 | 87.65 | 84.51 |
| | ✗ | 87.01 | 86.12 | 83.45 |

✓ means separate extraction ; ✗ means merge extraction

## V. CONCLUSION

This paper introduces a deep learning framework designed for detecting anomalous attacks within the Internet of Things (IoT), leveraging Long Short-Term Memory (LSTM) networks and an attention mechanism. It employs a feature extraction technique tailored for diverse variable types. The model's performance was benchmarked against baseline models on several datasets, including IoT-23, MQTT, IoT-NI, BoT-IoT, and MQTTset, with experimental outcomes demonstrating superior results. Additionally, comparative and ablation studies were performed to substantiate the efficacy of each module and the proposed methodologies. Subsequent research will concentrate on anomaly detection in imbalanced datasets.